\documentclass[sigconf,noacm]{acmart}
\setcopyright{none}
\renewcommand\footnotetextcopyrightpermission[1]{}
\acmConference[arXiv.org]{arXiv.org}{July 2026}{Boston, MA, USA}
\usepackage{etoolbox} % toggle
\usepackage[capitalize,noabbrev,sort]{cleveref} % cref
\usepackage{listings}
\usepackage{subcaption}
\usepackage{wrapfig}
\usepackage{listings}
\usepackage{url}
\usepackage{algorithm}
\usepackage{textcomp}
\usepackage{tikz}
\usepackage{amsmath}

\usepackage[export]{adjustbox} % makebox
\usepackage{multirow}
\usepackage{hhline}
\usepackage{pifont} % ding
\usepackage[inline]{enumitem}
\setlist{leftmargin=5.5mm,topsep=2mm}
\setlist[enumerate]{label=\textbf{(\arabic*)}}

\sloppypar

\newtoggle{comments}
\toggletrue{comments}
\newtoggle{cut}
\toggletrue{cut} % if this is active nothing will show
\newcommand{\ignore}[1]{}

\newcommand{\ldfs}{$\lambda$\textsc{FS}}
\newcommand{\sysname}{\textsc{SLFS}}

\newcommand{\proxy}{coordinator}
\newcommand{\proxies}{coordinators}
\newcommand{\Proxy}{Coordinator}
\newcommand{\Proxies}{Coordinators}

\newcommand{\dataf}{data function}

\newcommand{\Sdataf}{$f_{Data}$}
\newcommand{\Smetaf}{$f_{Meta}$}

\newcommand{\Sdmfs}{$f_{Data}/f_{Meta}$}
\newcommand{\Sdandmfs}{$f_{Data}$ and $f_{Meta}$}

\newcommand{\DMFs}{Data and Metadata Functions}
\newcommand{\dmfs}{data and metadata functions}

\usepackage{enumitem}
\newcommand{\itemparen}[1]{\begin{itemize}{#1}\end{itemize}}
\newcommand{\para}[1]{ %\vspace{0.5ex}
	\textbf{#1}
	}
\newcommand{\ballnumber}[1]{\tikz[baseline=(myanchor.base)] \node[circle,fill=.,inner sep=1pt] (myanchor) {\color{-.}\bfseries\footnotesize #1};}
\newcommand{\wballnumber}[1]{\tikz[baseline=(myanchor.base)] \node[circle,draw,inner sep=1pt] (myanchor) {\bfseries\footnotesize #1};}

\begin{document}

\title
[\sysname{}: a Flexible, Low-Cost Distributed File System Using Serverless Designs]
{\sysname{}: a Flexible, Low-Cost Distributed File System \\Using Serverless Designs}
%\subtitle{Research Paper \#15}

 \author{Cheng Hao (Ryan) Yang}
 \authornote{Both authors contributed equally to this research.}
 \affiliation{%
   \institution{Northeastern University}
   \streetaddress{360 Huntington Ave}
   \city{Boston}
   \state{MA}
   \postcode{02115}
   \country{USA}
 }
 
 \author{Paola Alsharabaty}
 \authornotemark[1]
 \affiliation{%
   \institution{Northeastern University}
   \streetaddress{360 Huntington Ave}
   \city{Boston}
   \state{MA}
   \postcode{02115}
   \country{USA}
 }
 
 \author{Soufiane Jounaid}
 \affiliation{%
   \institution{Northeastern University}
   \streetaddress{360 Huntington Ave}
   \city{Boston}
   \state{MA}
   \postcode{02115}
   \country{USA}
 }
 
 \author{Cristina Nita-Rotaru}
 \affiliation{%
   \institution{Northeastern University}
   \streetaddress{360 Huntington Ave}
   \city{Boston}
   \state{MA}
   \postcode{02115}
   \country{USA}
 }
 
 \author{Ji-Yong Shin}
 \affiliation{%
   \institution{Northeastern University}
   \streetaddress{360 Huntington Ave}
   \city{Boston}
   \state{MA}
   \postcode{02115}
   \country{USA}
 }
 
%\renewcommand{\shortauthors}{Yang et al.}

\begin{abstract}
Large-scale distributed file systems must provision resources for peak demand, yet file access patterns fluctuate significantly, leaving substantial capacity idle during off-peak periods. Existing scaling mechanisms operate at the granularity of entire servers and take minutes to hours, making them unable to track the rapid, fine-grained load variations that file systems commonly experience. Serverless computing, with its millisecond-granularity elasticity and pay-per-use pricing, offers a compelling alternative. We present \sysname{}, the first distributed file system built with serverless functions for both data and metadata operations. \sysname{} implements file services on top of key-value stores, keeping function operations simple and short, and introduces a novel multi-threaded, short-lived server design that overcomes the cold-start problem while maintaining low cost. A policy-enforcing coordinator efficiently maps files to function instances, scales the system elastically, and controls function lifetimes to balance performance and cost. \sysname{} can flexibly run on diverse storage backends---from cloud-native services like S3 to user-managed key-value stores---enabling configurable cost-performance trade-offs. Our evaluation shows that \sysname{} mitigates cold starts by 580$\times$ compared to the base serverless design and outperforms \ldfs{}, EFS, and Ceph at up to 63\%, 68\%, and 63\% lower cost, respectively.

\end{abstract}

\maketitle

%-------------------------------------------------------------------------------
\section{Introduction}
\label{sec:intro}
%-------------------------------------------------------------------------------

Large-scale distributed file systems are composed of hundreds of server nodes
where some nodes handle file operations while others focus on storing
data~\cite{gfs, ceph, oceanstore, pond, farsite, frangipani}.
In practice, file system load fluctuates significantly: not all files are
actively accessed at any given time, and access patterns shift unpredictably
across time of day, workload phase, and tenant activity.
Yet, to meet peak demand and maintain acceptable latency, operators must
provision resources for worst-case load, leaving substantial capacity idle
during off-peak periods. This over-provisioning translates directly into
wasted cost.
Existing distributed file systems offer scaling mechanisms, but these operate
at the granularity of entire servers or storage nodes---adding or removing a
node involves data migration, rebalancing, and reconfiguration that can take
minutes to hours~\cite{ceph, hdfs}. Such coarse-grained scaling cannot
track the rapid, fine-grained load variations that file systems commonly
experience.

Serverless computing offers a compelling alternative. In the serverless model,
users register function images and the framework dynamically launches,
auto-scales, and terminates function instances on demand, charging only for
the resources consumed during
execution~\cite{serverless-intro1, serverless-intro2, serverless-intro3}.
This model provides two key benefits for fluctuating workloads.
First, \emph{elasticity}: function instances can scale from zero to thousands
and back in seconds, matching demand at a granularity far finer than
server-level scaling.
Second, \emph{cost efficiency}: because the pricing model charges per
millisecond of execution, applications with idle intervals between bursts of
activity pay only for the time they are active, unlike VMs that incur cost
whether busy or idle.
These properties have already enabled serverless designs for data
analytics~\cite{caerus, llama}, machine learning~\cite{cirrus, serverlessml},
and video encoding~\cite{excamera}, hinting at the potential of serverless
computing to host a broader class of backend services.

Distributed file systems are natural candidates to benefit from these
properties. Most file operations are simple and short-lived---translating file
offsets to block addresses, reading or writing a few blocks---and many
applications issue file I/Os with intervals that leave the file system idle.
Capturing these operations in serverless functions can bring fine-grained
elasticity~\cite{lambdafs} and, when idle intervals are reflected in the
workload, significant cost savings compared to always-on server deployments.
For example, consider a multi-tenant platform where a media pipeline generates
bursts of small-file writes during upload windows then goes idle for hours,
a nightly analytics job reads terabytes once per day, and a web application
sporadically accesses configuration files. Today, these tenants either share an
over-provisioned file system---paying for peak capacity at all
times---or each deploys a dedicated cluster, multiplying operational overhead.
A serverless file system can instead scale each file's serving capacity
independently, from zero to many concurrent function instances, charging only
for the I/O actually performed.
\ldfs{}~\cite{lambdafs} explores this direction by extending
HopsFS~\cite{hopsfs} to delegate file metadata service to serverless
functions. However, \ldfs{} keeps function instances alive even during idle
periods to cache metadata, forgoing the cost benefits of the
serverless model.

In this paper, we propose, to our knowledge, the first serverless file system,
\sysname{}, which uses serverless functions to manage both file data and
metadata. Serverful storage systems exist for serverless
applications~\cite{pocket, boki, beldi, joe2020}, but \sysname{} uses
serverless functions as its main building blocks and can be used by both
serverless and non-serverless applications.
\ldfs{} only handles metadata using functions and is intended for
large file accesses like its predecessors, HopsFS and HDFS.
In contrast, \sysname{} is designed from scratch
for generic files, employs simple metadata designs suited to the serverless
environment, places both file data and metadata operations on serverless
functions, and adds mechanisms to efficiently utilize functions at low cost.

A key design decision in \sysname{} is to implement file operations on top of
key-value stores, keeping the computation within serverless functions simple
and short. Rather than managing complex on-disk structures such as
inodes and block bitmaps inside ephemeral functions, \sysname{} uses hash-based
mappings from file paths and block numbers to key-value pairs, similar to how
Ceph~\cite{ceph} uses hashing to map objects directly to storage nodes
without centralized metadata lookups. This design is a natural fit for the
serverless model: functions perform straightforward key-value gets and puts,
avoiding the heavyweight state and long-running logic that would conflict with
short function lifetimes. However, unlike Ceph, \sysname{} does not
manage the key-value backend itself; it delegates management complexity such as
replication and load balancing to the storage backend.
Thus, \sysname{} can run on any storage system with key-value APIs,
taking advantage of existing cloud key-value
services~\cite{awss3, azureblob, googlecs} as well as users' own key-value store
implementations~\cite{awsebs, azuredisk, googledisk}. The backend
choice can be diversified depending on performance, cost, management, and
other requirements.

The financial cost of operating \sysname{} is configurable and significantly
cheaper than using a cloud-native file system or \ldfs{}. Cloud providers
charge higher rates to store and access data in a file system than in a simple
key-value store or a block store, but there are still application demands for
file services. \sysname{} targets such applications, and our analysis of Azure
traces~\cite{azuretrace} under various \sysname{} configurations reveals that
the operational cost of \sysname{} is cheaper than that of cloud-native
file systems~(\cref{sec:cost}).
%% [CHANGED] Tightened: removed redundant clause, clarified the comparison chain
\ldfs{} reduces cost compared to open-source file systems such as
HDFS~\cite{hdfs} and HopsFS~\cite{hopsfs}, but \sysname{} reduces cost
further by dynamically controlling function lifetime while maintaining
high performance~(\cref{subsec:eval:cold,subsec:eval:cost}).

Despite these advantages, realizing a serverless file system is non-trivial
and entails unique challenges that stem from the serverless environment. Like
any serverless application, \sysname{} must deal with the cold start
problem~\cite{serverless-intro1}. Depending on the serverless platform,
starting a function from a cold state (i.e., starting from scratch) can take a
few seconds, and even from a warm state (i.e., starting from a paused sandbox
with a loaded function image) can take over 100s of
milliseconds~\cite{serverlessbench}. To mitigate cold starts and save
function invocation costs, \sysname{} introduces a novel function instance
design as a short-lived multi-threaded server. This design induces hot starts
(i.e., processing requests with a fully active function instance) and keeps
the instance busy even when waiting for an I/O response.

Supporting atomicity and isolation over weakly consistent backend storage
systems is another significant challenge in the serverless
environment~\cite{beldi, boki, ftshim}. \sysname{} dynamically assigns a
dedicated function instance for each file for isolation: that is, a file is
encapsulated in a function instance and is mutated by calls defined in the
instance, following an object-oriented design~\cite{lambdaobj}. This design also
facilitates a file cache within function instances. \sysname{}
leverages atomic commit features of backend key-value stores for consistent
updates and uses the two-phase commit protocol when the backend lacks native
atomic commit support.
\sysname{} provides single-file atomicity, and atomic multi-file operations 
such as rename.

Because most serverless frameworks do not fully support inbound networking of
function instances, mapping a file to a function and forwarding the file
request to the dedicated function is challenging. \sysname{} adds a \proxy{}
to control the request flow to the functions. The \proxy{} enforces
\sysname{} policies, mitigates cold starts, load-balances requests across
function instances, and scales the system.

We demonstrate through evaluation that \sysname{} is efficient, flexible,
scalable, and cost-effective. We compare \sysname{} against
InfiniCache~\cite{infinicache}, \ldfs{}~\cite{lambdafs}, AWS Elastic File
System (EFS)~\cite{awsefs}, and Ceph~\cite{ceph} and show that \sysname{}
delivers comparable or better performance at lower cost. \sysname{} takes
advantage of various storage backends while mitigating cold starts by
580$\times$ compared to default serverless configurations, improving
performance significantly, and reducing cost by up to
63\%, 68\%, and 63\% compared to \ldfs{}, EFS, and Ceph, respectively.

\sysname{} proposes a new way of constructing a distributed file system using
the serverless paradigm, offering a cost-effective alternative
to existing cloud file
services~\cite{awsefs, azurefile, googlefilestore, ceph}.
This paper makes the following contributions:
\itemparen{
    \item We propose the first serverless file system that uses
        serverless functions for handling both metadata and data.
    \item We present a novel short-lived-server-based function instance design,
        a \proxy{} design, and policies for the serverless file system to
        reduce cold starts, save cost, and scale. These designs are not limited
        to \sysname{} and can be applied to other serverless systems and
        applications.
    \item We implement \sysname{}, deploy it in the public cloud, and
        evaluate it under various scenarios to demonstrate its strengths and
        weaknesses against other systems.
}

\section{Background and Motivation}
\subsection{Cold Start Problem}
\label{subsec:coldstart}
Serverless computing allows the convenience of on-demand execution of functions
by launching the code and the runtime environment, usually in a container. The
trade-off for this approach is that if the function has not been executed for a
while, it takes a long time to fetch the image of the container and start 
the runtime environment. Mitigating the delays from this cold start
problem is an active research area. There are solutions such as creating a
lighter-weight sandbox~\cite{agache_firecracker_2020}, using snapshots of the
container to reduce the start-up
time~\cite{catalyzer,ustiugov_benchmarking_2021, seuss}, and predicting the task
invocation patterns~\cite{icebreaker, faascache} to keep the container in a warm
state (i.e., loaded but in a dormant state) for quick reuse. Some cloud
providers offer a service to keep the function instance warm (e.g., Azure
Durable functions and SnapStart~\cite{azuredf,snapstart}) at a higher price.
\sysname{} can benefit from these but proposes a more proactive
methodology and policy to keep the function hot while maintaining a low cost
(Section~\ref{sec:policy}).

%\subsection{Cost Analysis}
\subsection{Motivation and Cost Analysis}
\label{sec:cost}

\begin{figure}
	\centering
	\includegraphics{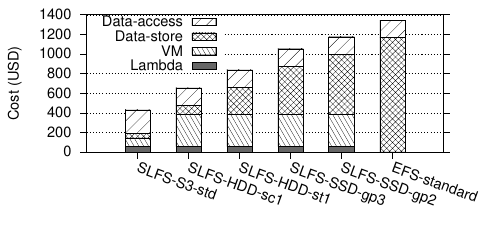}
	\vspace{-0.2in}
	\caption{\label{fig:awsprice} Cost comparison~\cite{awspricecalc} of \sysname{}
	under different configurations and a cloud native file system, EFS.}
\end{figure}

One of the main reasons users migrate their services to a serverless 
environment is the cost-effectiveness from the fine-grained resource management. 
While serverless computing in general is considered cheaper than serverful computing, 
this is not always true. The cloud vendors charge slightly higher 
unit price for the same amount of resource usage in serverless functions than in VMs. 
Thus, using serverless functions non-stop is not cost-effective and only the applications 
with some idle periods could benefit from serverless designs. Storage accesses in 
the cloud exhibit such idle patterns and we conduct a cost analysis. 

Figure~\ref{fig:awsprice} compares the cost~\cite{awspricecalc} of using a 
serverless file system (i.e., \sysname{}) over a key-value store 
(i.e., S3 standard) and block stores (i.e., EBS volumes: gp2 SSD, gp3 SSD, 
sc1 HDD, and st1 HDD) with a cloud provider's backend file service (i.e., EFS)  
in the Amazon cloud. We evaluate the cost 
based on the trace in the Azure function BLOB dataset~\cite{azuretrace}. The 
trace captures over 44.3 million data accesses from serverless applications for 
two weeks on 1.9 TB of data.

Analyzing this trace, we find that the median application's inter-quartile
I/O interval exceeds 2{,}000~seconds,
meaning that for more than half of the applications, a dedicated server
would sit idle for over 30~minutes between bursts of activity. During
these idle intervals, a serverless file system incurs zero function cost,
whereas a traditional file system consumes provisioned
resources. This workload characteristic is the key enabler of \sysname{}'s
cost advantage.

Serverless functions (i.e., Lambda) are 
necessary to implement file system APIs over EBS and S3, so function invocation 
costs are considered. We conservatively assume that each I/O request triggers 
the function to run for 300 milliseconds based on our measurement running
production workloads on \sysname{} (\cref{subsec:eval:cold}). 
EBS volumes require mountable virtual machines (VMs).
Although EBS volumes are internally replicated, we consider user-managed triple 
replication and add the cost of 9 VMs (t3.medium instances), each equipped with a 
687 GB EBS drive. % (sc1, st2, gp3, or gp2).
\sysname{} extends the serverless framework with a \proxy{}
(Section~\ref{subsec:proxy})
%; the \proxy{} can be part of the serverless 
%framework, but 
and we include the cost of running it as 3 separate VMs 
(t3.medium). EFS migrates a file that is not accessed for 
30 days to cold storage. However, all files in the Azure trace are accessed 
within 14 days so we only consider standard storage. For all configurations, 
both data storing and data accessing (i.e., reads and writes) costs are 
included, assuming that the applications are running in the same cloud region. 
The data access cost for \sysname{} includes the data transfer cost 
across different services. We project monthly costs on different storage settings.

The figure shows that using the cloud-provided file service is the most costly, 
whereas using more primitive or slower storage services costs less. 
Compared to EFS, serverless file systems using S3, st1-HDD-based EBS, and 
gp2-SSD-based EBS are 68\%, 37\%, and 12\% cheaper, respectively.  
The function invocation cost for \sysname{} is tiny in the 
figure; this justifies using serverless
functions in \sysname{} to host real workloads.
Overall, EFS charges a significantly higher rate for storing
data, while \sysname{} offers various cost-effective 
options. We further analyze the cost of \sysname{}, EFS, and other
systems relative to performance in \cref{subsec:eval:cost}.

\section{\sysname{} Design Overview}
In this section, we introduce the high-level design of \sysname{}.
Then, we dive into detailed designs of \sysname{} and policies to overcome 
unique challenges in the following sections.

\begin{figure}
	\includegraphics[page=2]{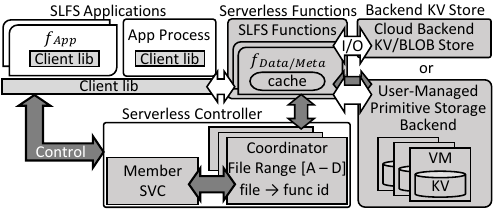}
	\vspace{-0.1in}
	\caption{High-level design of \sysname{} (shaded components).
	\label{fig:overview}}
\end{figure}

The design of \sysname{} is illustrated in Figure~\ref{fig:overview}.
\sysname{} is composed of the grey boxes in the figure. 
Main computations to support the file operations are designed as serverless
functions: the data function ($f_{Data}$) is responsible for accessing data 
files, whereas the metadata function ($f_{Meta}$) handles metadata 
operations. The file data is broken into
fixed-size blocks and stored as key-value pairs inside the backend storage
services. \sysname{} is designed to run on any cloud- or user-managed 
key-value stores. Similar to the Ceph~\cite{ceph} file system, \sysname{} 
uses the hash key of a file path and a block
number to map the block of the file to a key-value pair location. 
\Sdmfs{} operations are designed to be minimal for low-cost function
invocations. 

The \proxy{} orchestrates the function accesses. File requests first go
through the \proxy{}. The \proxy{} initializes the function instance 
dedicated to the file and establishes direct connections between the 
function and the client. The \proxy{} is the only 
serverful component in \sysname{}, which is similar to a proxy
in other serverless system designs~\cite{infinicache, lambdafs, infinistore}. 
We augment this design to avoid cold start, 
control concurrency, load-balance requests, and save \sysname{} costs.
The \proxy{} information is managed by the membership service natively 
in the serverless framework. 
The \proxy{} also scales up and down and the membership 
service in the serverless framework manages its information.

\subsection{Client Libraries and \sysname{} APIs}

The client library retrieves the \proxy{} information 
from the membership service and 
establishes connections to \Sdmfs{} with the help of 
the \proxy{} (\cref{subsec:optflow}). 

\sysname{} and the client library offer relaxed POSIX-compliant APIs 
similar to existing production and open-source distributed file systems~\cite{gfs, hdfs, awsefs}. 
The APIs include most file operations including open, close, read, write, 
delete, and rename and metadata operations.

The open call registers the file to the \proxy{}, and the \proxy{} assigns and 
launches \Sdmfs{} for the file.
The close call ensures that any outstanding I/O for the file is
completed and deregisters the file unless other users are accessing it. 
The file-to-function mapping can be dynamically changed by the \proxy{}, and 
long-term inactive files need remapping to a new function. % for reaccess.

Directories are implemented as files containing a list of files. 
Readdir reads the directory file and makedir creates a new directory file.
Files and directories can be renamed, but this entails an overhead of 
copying the file and directory to new locations because the hash
of the file path indexes files. 

\sysname{} guarantees per-file atomicity for multi-block operations,
serialized access for concurrent operations on the same file, and atomic
multi-file operations such as rename; we specify these guarantees precisely
in~\cref{subsec:consistency}.

\subsection{\DMFs{}}

A key design goal for \sysname{} functions is to keep operations simple and
fast so that function invocations remain short and inexpensive. To achieve
this, \sysname{} implements file operations on top of key-value stores using
hash-based indexing: read and write requests are translated into
block-granularity key-value accesses, where the hash of a file path and block
number directly maps to a key-value pair location without traversing any
sophisticated metadata structures. This design is critical in the serverless
environment. Serverless functions are short-lived and not well-suited for
caching inode-like metadata, and accessing a persistent backend store for such
metadata over the network can cause huge delays. Indeed, our early prototype
design with inodes was unusably slow and we had to switch to the hash-based
design. In contrast, \ldfs{}~\cite{lambdafs} caches file metadata in function
instances and extends the function lifetime to the maximum to overcome this
problem, but it incurs high costs and performs poorly when writing
metadata~(\cref{subsec:eval:meta}).

Building on this simple operation model, we propose a novel design for the
function instance: a \emph{short-lived concurrent server} that executes
simple tasks while keeping the function invocation cost low. Unlike regular
serverless functions that terminate after processing a single request, our
function instances run as multi-threaded servers that accept and process
multiple requests concurrently during their lifetime. The data function
(\Sdataf{}) handles file read and write requests by performing key-value gets
and puts based on the hash key. The metadata function (\Smetaf{}) is similar
but is dedicated to directory accesses; directory files are managed the same
as regular files, but \Smetaf{} implements directory-specific operations.
Because these operations are lightweight, they are well-suited to be processed
concurrently within the server-based function instance, keeping the instance
busy even when individual threads are waiting for backend I/O responses. The
lifetime of each instance is controlled by a policy that balances cost and
performance: the instance stays alive long enough to exploit temporal locality
and induce hot starts, but terminates when idle to avoid unnecessary
cost~(\cref{sec:policy}).

The \Sdataf{} and \Smetaf{} encapsulate file and directory instances as in
object-oriented (OO) style designs. Each function instance knows how to mutate
and access a file or directory, and \sysname{} assigns at most one dedicated
function instance per file or directory. Thus, accessing the function instance
becomes analogous to accessing a file/directory object. The OO design cleanly
isolates files/directories and facilitates concurrency
control~(\cref{subsec:consistency}). A similar OO function instance design was
proposed by LambdaObject~\cite{lambdaobj}, but \sysname{} further extends the
idea. First, read-only files can be mapped to multiple functions for
concurrent reads. Second, \sysname{} maps multiple files to one function
instance, since operations are the same across all files/directories. This
facilitates function instance reuse to avoid cold start and improves function
utilization. The load can also be dynamically distributed to different
functions to optimize performance and
cost~(\cref{subsec:coldstart,sec:policy}).

Leveraging the file-to-dedicated-function mapping, a strongly consistent
write-through LRU cache is kept in each function (\cref{subsec:optcache}). With the cache, the function
does not always have to access the backend store for reads. The cache is
designed to use excess memory in the function instance which would otherwise be
wasted.

\subsection{\Proxies{}}
\label{subsec:proxy}
The \proxy{} maps a file to a 
function instance. This mapping policy determines the performance and cost of
\sysname{} and keeps concurrent file accesses safe (\cref{sec:policy}). 

Upon opening a file, the \proxy{} checks if the file is 
already open. If not, the \proxy{} maps the file to an existing function or 
starts a new one to map the file. The mapping information is kept 
until no user accesses the file or the function terminates. 
Between the open and close of a file, different functions can service 
the file requests, but the \proxy{} ensures that at most one function 
instance serves the file at any given moment.

The \proxy{} can scale up or down depending on the load. \sysname{} uses 
consistent hashing like Chord~\cite{karger1997,stoica2001} to assign files to 
\proxies{}. However, unlike 
Chord, all \proxy{} information is kept in 
a centralized membership service in the serverless framework for consistent 
member updates and one-hop request delivery to the \proxy{}.

\begin{figure}
	\includegraphics{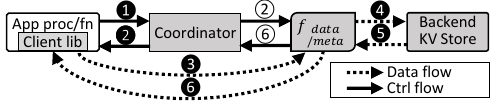}
%	\vspace{-0.1in}
	\caption{Separation of control and data flows.}
\label{fig:cdseparation}
\end{figure}

\subsection{Data and Control Flow Separation}
\label{subsec:optflow}

Although requests to \sysname{} start by contacting the \proxy{}, the 
\proxy{} is mainly responsible for handling the control flow, and the data flow 
bypasses the \proxy{}. 
We allow clients to directly communicate with 
\Sdataf{} and \Smetaf{} so as not to
overload the serverful \proxy{} (Figure~\ref{fig:cdseparation}). 
For \sysname{} operations, the \proxy{} tells the client about the function 
instance in charge of the file that the client is trying to access 
(\ballnumber{1} and \ballnumber{2}). 
If a new function instance is needed to process the client request, 
the \proxy{} starts a new function instance (\wballnumber{2}) before returning to the client.
Next, the client directly contacts the 
function instance for the file access (\ballnumber{3}) and the instance performs
corresponding key-value accesses (\ballnumber{4} and \ballnumber{5}).
The function instance returns the result of the file access to the client (\ballnumber{6})
and also notifies the \proxy{} about its status (e.g., number of queued requests)
so that the \proxy{} can make policy-based decisions for \sysname{} (\wballnumber{6}). 

While open source serverless frameworks %be configured to 
allow inbound connection to function instances, some commercial platforms 
prohibit this. As a workaround, one can use TCP/NAT hole punching~\cite{ford2005a, wukong}. % (this is known to work in AWS). 

\subsection{Storage Backend}

\sysname{} is designed to work with key-value storage backends. The condition
for safety (see Section~\ref{subsec:consistency}) is that the key-value store should provide 
read-my-writes %~\footnote{A user sees the same or later version for the object 
%that the user wrote.} 
and bounded staleness, %~\footnote{Data can be stale by at most time $t$.} 
or stronger guarantees~\cite{pileus}: 
users must always see their own updates, and an update must become visible 
to everyone after a bounded time. % $t$. 
Most cloud storage systems, such as S3~\cite{awss3}, fulfill this requirement.
To further demonstrate a more general use case of \sysname{}, we also consider even more primitive storage 
building blocks based on standalone VM-based key-value storage servers
(Section~\ref{sec:backend}).

\section{Detailed Designs}
\label{sec:detailed-designs}
\begin{figure}
\centering
	\includegraphics[page=1]{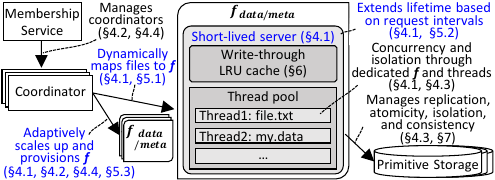}
%	\vspace{-0.1in}
    \caption{Relationship between \sysname{} components and detailed designs.
	Dynamic policy designs are in blue.
	\label{fig:detail}}
\end{figure}

%In addition to the base design of \sysname{},
We describe the detailed \sysname{} designs to 
answer the following questions for realizing a
practical serverless file system:
%answer these questions.
%There are questions to address for an efficient 
%serverless file system designs:
1) how to avoid cold starts;
2) how to lower the financial cost;
3) how to scale the components; 
4) how to maintain consistency % requirements; 
and
5) how to handle failures.
Figure~\ref{fig:detail} illustrates the 
detailed relationships and roles of \sysname{} components.

\subsection{Avoiding Cold and Even Warm Starts} 
\label{subsec:coldstart}

Cold-starting a function instance can take a few seconds, and even warm-starting 
an instance can take 100s of milliseconds~\cite{serverlessbench}. 
The recent AWS SnapStart feature reduces the start-up time to a few
milliseconds~\cite{snapstart} with a warm snapshot of a function instance, but 
for fast file accesses, it is best to even avoid the warm start. 
We propose a novel short-lived multi-threaded server-based function design 
for \sysname{} to induce a hot start.

\para{Function Instances as Short-Lived Servers.} We design \Sdmfs{} as 
short-lived servers to reduce the file access latency. 
Unlike regular serverless function instances that terminate right after processing a request,
\sysname{} functions wait for new incoming requests for a short period leveraging
the temporal locality of file accesses. 
If there is another request to the same file within a short interval, 
the instance can be immediately reused; this leads to a hot start. 
%Since files generally exhibit temporal locality, this design is 
%effective. 
The limitations are that cold or warm starts are 
unavoidable under infrequent I/Os, and extra costs could incur 
for idling. Thus, the function lifetime policy is designed 
to keep the cost low while inducing hot starts
(Section~\ref{subsec:policy-function-lifetime}). 

\para{Multiple Files Per Function Instance.} If a function instance is only
mapped to a single file, the frequency of the instance reuse will be 
low, and new file accesses will entail cold starts. To better utilize a 
function, we assign multiple files per function 
(Section~\ref{subsec:policy-file-to-function}) and make the instance
multi-threaded. For security reasons, however,
only the files of the same application/user share a function.
%files share the 
%function only when they are accessed by the same application/user.

Concurrently handling different file requests in the same function instance
saves function invocation costs: e.g., cold-starting one instance
prepares multiple file requests to be processed immediately. Without 
concurrent request processing, functions could idle while waiting for I/O 
responses from the backend storage, but our design keeps the function busy 
and reduces the overall function execution time. 
As long as the user frequently accesses any file the utilization of 
the function instance can increase, and cold and warm starts can be avoided. 
Thus, for example, scanning through a list of files or recursively 
reading file lists in directories (e.g., find or ls -R)
is handled efficiently using only a few functions.
 
\para{Provisioning Replacement Functions.}
Even though the \Sdmfs{} can handle multiple requests for different files without
having to terminate after processing a request, 
%this does not mean  
%the function can live forever. 
%Function instances have a limited lifetime 
it has a limited lifetime set by the serverless 
framework~\cite{wang_peeking_2018, shahrad_serverless_2020}. 
In case there is a need to process I/O requests beyond the function's
lifespan, each instance keeps track of its duration. 
The function instance notifies the \proxy{} when it is close to the 
termination time set by the framework. Then the \proxy{} launches a replacement 
function early enough to hide the cold start. Later, the old function 
terminates with notifications to the \proxy{} and connected clients, and
the \proxy{} directs requests to the new function. 

\subsection{Scaling Up and Down}

For any serverless application, scaling up and down is important 
for both performance and cost. The scalability of \sysname{} is only 
bounded by the backend storage that \sysname{} relies on.

\para{Scaling \Proxies{}.} 
The \proxy{} scales up and down depending on the user load. When
the number of open files a \proxy{} manages increases, the \proxy{} spawns 
a new \proxy{} server to partition the range of files it manages. The existing
\proxy{} reorganizes the file to function mapping, function instances
that manage files for the new \proxy{} connect to the new \proxy{},
and the new \proxy{} updates its file to function mapping. The new \proxy{} 
becomes active after its information is registered with the membership service. 
Due to the consensus-based membership service (i.e., Zookeeper~\cite{zookeeper})
the function instance mapping and the hand-off of a partition take 
place consistently and then clients of \sysname{} are notified of 
the changes through the client library.

When the load on the \proxies{} goes below a watermark, \sysname{}
merges adjacent \proxy{} nodes through the membership service similar to the
scaling up procedure.  
Overall, scaling up a \proxy{} is much more costly than scaling a function
instance, so \sysname{} triggers the merger when the load is kept below the
watermark for an extended period.

\para{Scaling \DMFs{}.} Because we map multiple files to a function
instance, \sysname{} should opportunely launch new instances as the 
load increases. \sysname{} monitors the request queue length and starts new 
functions if the queue length exceeds a threshold. It considers the cold 
start latency and proactively provisions function instances for the best 
performance (Section~\ref{subsec:policy-new-function}).
%Due to the short-lived nature of serverless functions, 
%\sysname{} does not actively scale down the functions. 
For scaling down, the function lifetime policy (Section~\ref{subsec:policy-function-lifetime}) 
terminates the function without creating a replacement and adjusts to the load.

\subsection{Maintaining Consistency}
\label{subsec:consistency}

\para{Guarantees.}
Before describing the mechanisms, we summarize the guarantees \sysname{}
provides to clients: (1)~reads and writes to a file spanning multiple 
blocks are atomic; (2)~concurrent accesses to the same file are serialized; 
%(3)~directory updates never create dangling references;
(3)~multi-file operations such as rename are atomic;
and (4)~completed writes are durable.

\para{File Consistency.} 
\sysname{} uses a mix of soft-update~\cite{softupdate1,softupdate2}
and lock-based two-phase commit~\cite{2pc, sinfonia} to guarantee file consistency. 
\sysname{} does not employ an explicit index such as the inode, and
locations of file blocks are derivable by a hash function. The only
persistent index-like metadata is the directory file, and files synchronize 
through the directory. For example, the directory is updated after a file 
is created, and the file is deleted after the directory entry of the file 
is removed. If the operation fails halfway, there could be orphaned files, 
which can later be garbage collected, but inconsistent states are not revealed
to the user. Thus, the directory information acts as the ground truth for 
what is in the file system. Note that concurrent updates to the directory file 
are safe because a dedicated \Smetaf{} serializes the concurrent operations
for each directory file. For multi-directory/file updates (e.g., rename), one of 
the functions registers the transaction with the \proxy{} and works as a transaction
coordinator to drive the two-phase commit protocol~\cite{2pc} over the functions
that own the relevant files.

For atomic reads and writes to multiple blocks of the same file, we leverage the 
transactions or atomicity features~\cite{dynamodb-tx, s3-atomic} supported by
the backend key-value stores. In case no such features are supported, \sysname{}
uses the two-phase commit to atomically read or write multiple blocks. This feature
is activated when \sysname{} runs on top of our custom backend key-value stores (\cref{sec:backend}).
%of a file from the backend key-value store.

\para{Replicated Consistency.}
Depending on the backend key-value store and its replicated consistency 
semantics, stale versions of a file can be exposed. 
However, as long as the backend key-value store supports read-my-writes or stronger
semantics~\cite{pileus, sessionev}, 
\sysname{} can service the up-to-date state of all files it manages and 
preserve consistency across different key-value pairs. 
Note that the weak consistency guarantees such as read-my-writes are often defined 
per user basis but the user of the key-value store in our context is the entire 
\sysname{} which directly interacts with the key-value store. Such consistency guarantees
for the entire \sysname{} extend strongly to individual end users. For example,
under read-my-writes, an end user's completed write becomes immediately visible
to another end user even though they are different users, because \sysname{} as a 
single user of the key-value store reads and writes on behalf of all end users. 
All backend stores used for \sysname{} in this paper, including Amazon S3 
and our own key-value store implementation, satisfy this requirement.

\subsection{Handling Failures}

Similar to most serverless applications, \sysname{} functions are 
stateless, and critical states are maintained in the persistent backend store. 
We assume the backend store is durable (e.g., due to triple replication) and 
the membership service is highly available (e.g., backed by Paxos-like 
protocols~\cite{paxos, zookeeper}). 
With these assumptions, 
\sysname{} can tolerate the failure of \dmfs{} and \proxies{} and maintain
consistency. 

\para{Failures of \DMFs{}.} \Sdandmfs{} are stateless worker functions 
managed by the \proxy{}. If they fail, the \proxy{} can detect
the failure, reassign the files of the failed function to other functions, 
and redirect the file requests accordingly. 
If the \proxy{} is unsure whether a function instance has failed or not, it 
can ask the serverless framework to terminate the suspected instance 
and switch to a new one. 
Any queued requests in the failed function will be returned as failed I/O or
timeout, and users are responsible for re-executing them.  
\sysname{} guarantees the atomicity of file I/O requests, so function failures 
do not cause inconsistencies. Still, \sysname{} needs to abort or complete 
incomplete I/O operations involving two-phase commits which follows
the two-phase commit recovery protocol of checking transaction intention records 
left behind in the backend key-value stores. 

\para{Failures of \Proxies{}.} The \proxy{} manages states that outlive the
\Sdandmfs{}, and handling its failure is more complicated than
that of function instances. 
%Because \proxies{} are registered to the membership service, 
\sysname{} uses the heartbeat mechanism to detect failures using the membership 
service where the \proxies{} are registered. If the failure is detected, a replacement 
\proxy{} is registered, and an epoch number is incremented. 
Functions of the failed \proxy{} can continue processing requests 
but they terminate when they cannot communicate with the 
\proxy{}; the new \proxy{} starts operating after all old
functions terminate. 
%Once the new \proxy{} can ensure that the old 
%unctions are no longer alive, it starts processing requests. 

\para{Recovery of In-Flight Transactions.}
Multi-file transactions in \sysname{} (e.g., rename) use two-phase commit,
where one function instance acts as the transaction driver. To enable
recovery under failures, the driver writes a transaction intention
record containing the participant list, transaction ID, and current
state to a well-known key prefix in the backend key-value store
before sending prepare messages to participants. Each participant
records its vote (commit or abort) in the backend store as well.

This design enables recovery even under simultaneous failure of the
\proxy{} and the driving function. When a new \proxy{} starts after a
failure, it scans the backend store for pending intention records. For each
pending transaction, the new \proxy{} determines the outcome: if all
participants have voted to commit and a commit record exists, the
transaction is completed by ensuring all participants apply the update;
otherwise, the transaction is aborted and participants roll back. This
follows the standard two-phase commit recovery protocol~\cite{2pc}.
The overhead of this mechanism is one additional key-value write (the
intention record) per multi-file transaction. In our workloads, multi-file
transactions are rare, so the overhead is negligible.

\section{\sysname{} Policies}
\label{sec:policy}

\sysname{} maintains internal policies to make decisions for 1) how to map
a file to a function instance, 2) how long the function instance should 
wait for new requests, and 3) when to launch new function instances. 
The policies affect and consider the performance and the cost of
running \sysname{}.

\subsection{File-to-Function Mapping}
\label{subsec:policy-file-to-function}

\sysname{} file requests go through the \proxy{}, so the \proxy{} keeps 
track of how many files are opened and how many I/O requests for each file are 
queued in the function. We have explored random and greedy policies to map
a file to a function. The greedy policy assigns files to the function with 
the shortest request queue size. Based on an analysis using the 
Azure serverless I/O trace, we find that the greedy policy is simple yet 
works the best in general and use it for \sysname{}. Even after the initial  
assignment, the file-to-function mapping can change dynamically. 
The \proxy{} periodically monitors the load on each file and remaps the heavily
accessed files to the least loaded or a new function instance. 

\subsection{Function Lifetime}
\label{subsec:policy-function-lifetime}

An extended function lifetime trades off function running cost for low latency. 
Keeping the function hot all the time as in \ldfs{}~\cite{lambdafs} yields the best 
performance, but this nullifies the purpose of the demand-based serverless 
design. While other systems mathematically 
predict the load to keep the function instance \emph{warm}~\cite{shahrad_serverless_2020, icebreaker, faascache}, 
\sysname{} maintains the functions \emph{hot} for 
a shorter period with higher sensitivity.

Our policy is, therefore, simpler and uses 
a moving average of request arrival intervals of the last $N$ requests for each 
function. The function instance waits for $T_{ext}$, the average 
interval with an added error margin $\epsilon$, after it processes all queued 
requests. However, if the request arrival intervals are very short,
%or requests arrive concurrently, 
reflecting close-to-zero intervals to $T_{ext}$ does not make sense. 
Thus, we only consider intervals greater than a threshold 
$T_{min}$. To reduce the effect of outliers, 
we provide a safety threshold using the coefficient of variation 
(standard deviation / mean). If the coefficient is too high, we revert
to a default $T_{default}$ until the spread 
%of request intervals 
gets smaller.

\subsection{Launching New Functions}
\label{subsec:policy-new-function}

Because the serverless pricing model charges for the duration of a function, the
price for serially invoking the same number of functions is the same as concurrently 
invoking them. 
\sysname{} functions internally
parallelize and pipeline tasks and extend their own lifetime, so opportunely launching
new functions is critical to improve performance and save costs. 

\sysname{} starts a new function instance 1) when a
function instance is about to be terminated by the serverless framework
due to the maximum run time limit, and
2) when there are not enough function instances to handle the
load. In the former case, the \sysname{} \proxy{} starts a replacement
function while the terminating function is running to switch and hide the cold
start latency. For the latter case, \sysname{} profiles the
steady-state performance of \Sdandmfs{} and measures the maximum load that the
functions can handle without sacrificing performance. The \proxy{} monitors the
load based on the function's feedback and launches new functions. 
If a burst of requests arrives, the \proxy{} starts multiple functions proportional to 
the load in parallel. In this case, cold or warm starts are
unavoidable. However, the multi-threaded design of the function handles
the load more efficiently with a smaller number of instances than regular serverless
applications that start an instance per request.

\begin{figure}
	\includegraphics{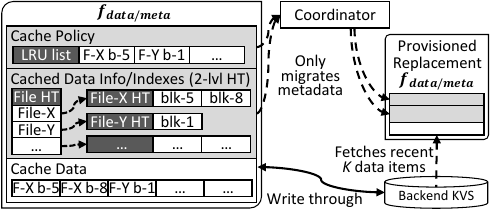}
%	\vspace{-0.1in}
	\caption{\sysname{} cache design and cache migration (dotted arrows) 
	to replacement function instances.}
	\label{fig:cachedesign}
\end{figure}

\section{Caching}
\label{subsec:optcache}
We implement a write-through least recently used (LRU) block cache inside 
each function to increase the performance and memory utilization
(Figure~\ref{fig:cachedesign}). 
\sysname{} places the cache inside the function instance for three reasons. 
1) \sysname{} function instances usually underutilize allocated memory, so  
the cache comes almost for free. 2) \sysname{} files are mapped to dedicated 
function instances, so there is little cache contention, and 3) a single 
copy of cached data can serve all requests to a specific file without 
cache consistency concerns. The write-through approach by the
dedicated function retains durability and consistency. Two levels of 
hashing using file names and block numbers index the cached data.

A unique challenge in the serverless environment is that a function instance
with cached data frequently terminates. Before the instance terminates, it sends
the information about the cached blocks (not the data) of each file (2nd-level
hash table) to the \proxy{}. When the \proxy{} remaps the file to a new function 
instance, this cache information is sent to the new instance. Then, the new instance
prefetches recently accessed $K$ blocks of the file from the backend key-value 
store into its cache for continued caching. 
The cached block information that is temporarily stored in 
the \proxy{} is small. However, the \proxy{} keeps a cap and evicts the 
block information at file granularity based on the LRU policy.
Note that the typical cache information size per function is only a few hundred 
bytes, and the cache can operate without the information
transfer.

\section{Primitive Storage Backends (PSB)}
\label{sec:backend}

Compared to a cloud-provided file service, \sysname{} users can 
flexibly employ different backend storage systems while 
maintaining the serverless elasticity. As a demonstration, 
we run \sysname{} over a primitive storage backend (PSB).

We configure a set of VM nodes that run LevelDB on a mounted virtual block
device. Each node runs a server layer that interfaces with the 
LevelDB~\cite{leveldb} for key-value accesses. These storage nodes run 
independently, but \sysname{} ties them together as a distributed file system. 

\sysname{} controls where each key-value pair corresponding to a 
block of a file is stored. In this design, individual storage nodes are managed 
as a Chord-like consistent hash ring. Each storage node stores key-value 
pairs that fall under a specific key range, and the hash value of the file block 
directly maps to a key-value pair in one of the storage nodes. 
We made the block placement configurable such that the blocks from the same 
file are stored either on the same node or on different nodes.
%\sysname{} manages how the data is kept reliably and consistently. 
\sysname{} controls the number of file replicas and how each file is replicated. 
We use the client-driven chain-replication scheme~\cite{chain-replication, corfu} 
to ensure that file blocks are replicated consistently over multiple nodes. 
This ordered replication scheme makes it easy to recover from potential failures. 
\sysname{} replicates updates to different files concurrently, and blocks of the 
same file are replicated in a pipeline for performance.
These schemes for the PSB are implemented as an added feature of  
\Sdandmfs{}. Note that the PSB design is an example, and users can have their custom 
design or use existing key-value stores.
%We show in the evaluation that 
%\sysname{} runs faster at a lower financial cost on top of the PSB 
%compared to a full-featured distributed file system. 

\begin{table}
%\footnotesize
\ttfamily
\resizebox{3.4in}{!}{%
\begin{tabular}{l|l r r r c r}
\hline 
&Name  & \# files & \# I/Os & Accessed & R:W & Interval\\
\hline 
r1&	01qqaww4    &   2       &   732.7 K &   18.8 MB     &   100:0   &   0.09 s   \\
r2&	7xjpt3h0    &   87.5 K  &   86.7 K  &   1.7 GB      &   100:0   &   11.66 s  \\ 
r3&	66wi1vut	&	2.9 K	&	2.9 K	&	113.1 MB	&	100:0	& 875.12 s \\
\hline 
w1&	0insggn2    &   4.1 M   &   8.2 M   &   56.2 GB     &   0:100   &   0.08 s   \\
w2&	kla1jqcy	&	149.6 K	&	149.6 K	&	216.1 MB    &	0:100	&	28.87 s  \\
w3&	u1q65cm8	&	1.3 K	&	1.3 K	&	158.0 MB	&	0:100	&	139.26 s \\
\hline 
b1&	7agnu66h    &   3.7 K   &   6.0 K   &   1.7 GB     &   20:80   &   1.00 s   \\
b2&	j03uywqk    &   1.8 K   &   2.8 K   &   7.5 MB      &   64:36   &   12.11 s   \\
b3&	xlxky5ax    &   24      &   6.5 K   &   1.1 GB      &   50:50   &   117.89 s \\
\hline 
\end{tabular}%
}
\caption{Selected applications from the Azure BLOB trace and their characteristics.}
\label{tbl:trace}           
%	\vspace{-0.1in}
\end{table}

\section{Evaluation}
In this section, we evaluate the basic performance and scalability of 
\sysname{}, how well \sysname{} avoids cold and warm starts, and
the performance and cost of \sysname{} against other systems. 

\para{Implementation and Basic Configurations.}
We implement \sysname{} on OpenWhisk~\cite{openwhisk} running with Docker
containers~\cite{docker}. 
We use the Zookeeper included in the OpenWhisk package as 
the \sysname{} membership service. All of the \sysname{} components, including
the user library, \proxy{}, \Sdataf{}, \Smetaf{}, and PSB, are implemented
in 12.6K lines of C++ code. \Sdandmfs{} are integrated as a single Docker image for easy reuse of the 
function instance.

We evaluate \sysname{} using 32 VM nodes 
with 2 vCPUs and 4GB DRAM. We use 7 VMs to host components related to the 
serverless framework, which include the OpenWhisk controller, Zookeeper, and 
\proxies{}, 16 VMs to run OpenWhisk workers and Docker-based function 
instances, and 9 VMs to set up the backend storage. We use function
instances with 256 MB memory and run PSB or Cassandra key-value store~\cite{cassandra} 
on the 9 storage VMs. 
For both PSB and Cassandra setups, all VMs play the same role and store key-value 
pairs. Both setups maintain three data copies across different nodes by default. 
Unless mentioned otherwise, \sysname{} runs on the PSB with 3 \proxy{} nodes
and is evaluated on a public research cloud. 

\para{Baselines.}
We use open-sourced \ldfs{}~\cite{ldfscode} to evaluate the cold-start frequency 
and cost for running the Azure workload (\cref{subsec:eval:cold}) and the 
performance of metadata operations (\cref{subsec:eval:meta}). 
We allocate the same amount of hardware resources to \ldfs{} as \sysname{} for 
the serverless framework, function instances, and storage nodes, respectively, 
but the main differences are:
	1) individual \ldfs{} function instances use a larger amount of 
	vCPU and RAM (i.e., 2 GB minimum) due to its inherent design and 
	2) the storage resources are divided into metadata and data 
	stores where the metadata is stored in MySQL Cluster NDB~\cite{mysqlndb} and
	the data blocks are stored as files in each node's local file system. 
\ldfs{} is the closest system to \sysname{}, but unfortunately, \ldfs{} only focuses 
on improving metadata operations and does not support data operations well.
Through personal communications with the \ldfs{} authors, we have confirmed that  
all experiments in the \ldfs{} paper do not include data operations and
persistent data access considerably slows down \ldfs{} 
(e.g., 21$\times$ slower than \sysname{} in \cref{fig:ldfs-read}). 
Thus, we do not use \ldfs{} for heavy performance evaluation in \cref{subsec:e2e}. 

We use the Ceph file system~\cite{ceph} and AWS EFS~\cite{awsefs} to 
compare the performance and cost against \sysname{} (\cref{subsec:e2e,subsec:eval:cost})
While EFS has little to configure, Ceph requires exploring
different configurations. Based on the same storage setting of 9 VMs as 
\sysname{} we experimented with different Ceph settings to optimize for 
performance. For Ceph at this scale, rather than distributing monitor (MON) and 
metadata servers (MDS) to separate VMs we found that collocating them with 
object storage daemons (OSD) within the same VM yielded the best performance. 
We used 225 placement groups and collocated 2 MONs and 3 MDSes within
9 VMs and the Ceph balancer tool returned a good score for this configuration. 

Finally, we compare \sysname{} against InfiniCache~\cite{infinicache} to evaluate the effectiveness of our caching scheme 
(\cref{subsec:eval:cache}).

\para{Workloads.}% Azure Traces and Trace Emulators.}
We use the Azure function BLOB dataset~\cite{faastcache, azuretrace}, 
YCSB~\cite{ycsb} benchmark, IOzone file system benchmark~\cite{iozone},  
	a uniform random workload, and \ldfs{} benchmark~\cite{lambdafsbench}
	and report the mean of at least three experiments. 

The Azure dataset contains 14-day traces of 855 applications. We analyze the
trace to find representative applications that access files. We collect 
the I/O intervals for each application trace and take the inter-quartile range 
(IQR) as a descriptive statistic. We use the IQR because it is not as sensitive 
to outliers and accurately captures the most probable range where a time 
interval can fall. The IQR intervals of applications above the 45th \%tile 
exhibit over 2,000 seconds, making them uninteresting for our evaluation. 
Thus, we focus on the applications below the 45th \%tile and categorize 
applications into three groups: 1) 0-15th \%tile, 2) 15-30th \%tile, and 3) 
30-45th \%tile. Within each group, we choose the applications with 
read-only (r), write-only (w), and balanced read-write ratios (b) with a large 
amount of I/O requests and byte accesses. We name each application by the 
read-write ratios and group names (e.g., r1 is a read-only workload from 
the 0-15th \%tile group). The selected applications and their characteristics 
are in Table~\ref{tbl:trace}.  

We built an emulator to replay the trace in two modes:
one that issues requests following the I/O interval in the trace 
%at the same speed as the original trace 
and the other that issues I/O requests 
sequentially as quickly as possible. We use the former to evaluate the 
cold/warm/hot start and caching effects and the latter for  
performance. 

\subsection{Basic Performance}
\begin{figure}
    \centering
\includegraphics{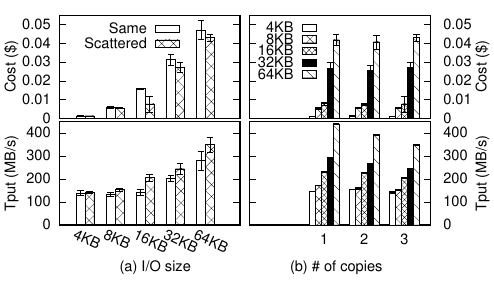}
	\vspace{-0.2in}
    \caption{Different block placement and number of replicas.}
    \label{fig:basic-vs-stripe} 
\end{figure}

When using the PSB, \sysname{} supports placing the blocks from the same file on
the same PSB node or scattering them across different nodes. Also, \sysname{} 
can configure the number of replicas that users maintain.
We evaluate the performance and cost of these settings using 512
concurrent clients issuing 2 million requests that span multiple blocks using
the YCSB-A workload (Zipfian distribution of 50:50 reads and writes). 

Figure~\ref{fig:basic-vs-stripe}a shows that regardless of the I/O size, the
throughput of scattering a file's blocks to different PSB nodes scales better and 
achieves higher throughput at a lower cost (i.e., the duration of all individual functions
converted into the AWS Lambda cost~\cite{lambda-price}). The \dataf{}
can access different PSB nodes much faster in parallel in contrast to accessing
blocks in a single PSB node. Thus, we choose the scatter data
layout for the remainder of the evaluation.

Next, we experiment with different numbers of file copies from 1 to 3 
(Figure~\ref{fig:basic-vs-stripe}b). As expected, creating more copies requires extra 
bandwidth and cost. Still, the throughput decrease and the cost increase 
between having 1 and 3 copies are only 12\% and 1\% on average, respectively.
Thus, we use 3 copies for all other experiments, but this experiment
shows that \sysname{} can trade off availability for cost and performance. 

\subsection{Elasticity and Scalability}

\begin{figure}
    \centering
\includegraphics{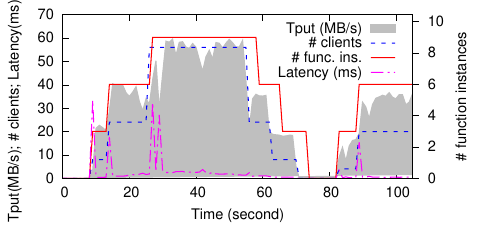}
	\vspace{-0.1in}
    \caption{Scaling of function instances depending on the load. The number of
	function instances closely follows the load. Latency spikes show 
	cold-starting function instances.}
    \label{fig:scaling} 
\end{figure}

\para{Scaling Function Instances Up and Down.}
To evaluate how well \sysname{} scales up and down elastically,
we dynamically change the amount of uniform random I/O loads with a 4KB request
size on \sysname{} by controlling the number of concurrent clients. 
Figure~\ref{fig:scaling} shows that the \sysname{} policy effectively scales  
the number of function instances to match the increasing and decreasing load
trend. Function instances start from a cold state at the
beginning of the execution so the latency spikes when new instances are added
(10, 15, and 30 second marks). Still, the number of function instances
closely and proportionally reflects the number of concurrent clients. The
function instances stay up until slightly after the client load decreases, but
they quickly adjust to the decreased load (
55-80 second marks). When scaling back up (83 second mark), the 
latency does not spike much due to warm instances. This experiment
demonstrates that \sysname{} scales up and down (even to zero at 75-83
second marks) quickly depending on the load and satisfies the serverless
elasticity while maintaining the server-based function instance design.

\begin{figure}
    \centering
    \includegraphics{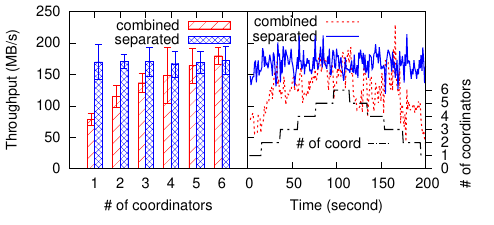}
	\vspace{-0.1in}
    \caption{Performance under a fixed number of \proxies{} (left) and 
		dynamically scaling \proxies{} (right) with and without the data flow 
		separation.}
    \label{fig:proxyscaling} 
\end{figure}

\para{\Proxy{} Scaling and Flow Separations.}
We manage a serverful
implementation of \proxies{} and measure the performance under different numbers
of \proxies{}. As a comparison, we use a baseline without the data flow 
separation introduced in Section~\ref{subsec:optflow} (Figure~\ref{fig:proxyscaling}). 
In the baseline, all data goes through the \proxy{}.
We measure the performance under the uniform random workload with a fixed 
number of \proxies{} (left) and while the \proxy{} scales up and down 
dynamically (right). 
Without the data flow separation, the \proxy{} can become the bottleneck 
as the data passes through it. When 
clients bypass the \proxy{} and directly interact with the functions, 
having a single \proxy{} can saturate the backend storage.
The dynamic scaling of the \proxy{} demonstrates that \proxies{} can smoothly
hand off files they are managing to different \proxies{} without disrupting 
the system execution.

\subsection{Caching Comparison with InfiniCache}
\label{subsec:eval:cache}
To evaluate the effectiveness of the \sysname{} caching scheme, we compare against 
our implementation of the InfiniCache~\cite{infinicache} design, which we place on a 
cacheless \sysname{}. 
Our InfiniCache baseline runs multiple long-running function instances to host the 
maximum amount of cache memory that \sysname{} uses during runtime. The 
instances cache a disjoint set of data in memory. 
When an instance reaches its max lifetime (5 minutes in OpenWhisk),
the instance terminates, and the cached data is discarded. The \proxy{} serves 
the role of InfiniCache proxy so the network hops for read/write 
operations stay the same as \sysname{}.
\begin{figure}
    \centering
    	\includegraphics{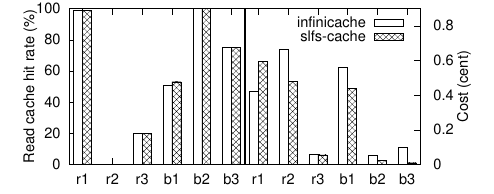}
	\vspace{-0.1in}
	\caption{\sysname{} cache and InfiniCache. \sysname{} cache performs comparably at a cheaper cost.}
    \label{fig:cache} 
\end{figure}

We run 1-hour sub-traces of the Azure trace that match the IQR interval with 
the full trace and replay them while observing the trace timestamp. Because 
\sysname{} uses the write-through policy, we only evaluate the traces that
include read requests (Figure~\ref{fig:cache}). 
The \sysname{} cache hit rate is comparable to InfiniCache. However, because 
\sysname{} passes the recently accessed data information to new instances, 
it exhibits a better hit rate when InfiniCache loses a chunk of cached data due 
to an instance termination (i.e., ``b1'').

Because InfiniCache requires running separate cache instances, the cost of 
running InfiniCache with \sysname{} is generally more expensive. ``r1'' is an 
exception as it consists of short-interval read-only requests with nearly 
100\% cache hit rates. Requests are mostly served from the cache without 
touching the backend store, and InfiniCache leverages the hot start supported by 
OpenWhisk. However, \sysname{} extends
the function lifetime to inject hot starts and adds costs. All data in ``r2'' is read once, 
resulting in 0\% cache hit rate. 

To quantify the performance improvement from introducing the cache,
we measure the end-to-end run time of \sysname{} with and without the cache.
We replay the r1-3 and b1-3 traces as quickly as possible and run YCSB-A and 
IOZone 50/50 random read/rewrite benchmarks similar to experiments 
in \cref{subsec:e2e}. On average, the cache improved performance 
by 22.68\% across all workloads.

\subsection{Cold, Warm, and Hot Starts}
\label{subsec:eval:cold}
We evaluate how \sysname{} reuses the server-based function instances using the
policies to prevent cold and even warm starts. Using the Azure trace, we measure
the frequency of cold, warm, and hot starts in \sysname{}.  We ran the experiment for
12 hours for each trace.  We select the 12-hour range that best represents the
overall request arrival interval IQR of the full trace and replay it while
observing the trace timestamp. For the baselines we use  
the default serverless function design and \ldfs{}.
%, we assume a straw man 
%\sysname{} design with a regular serverless function usage pattern: 
The default serverless function design follows a regular serverless function usage pattern: 
it terminates a function immediately after processing 
one request, and an independent function processes new requests. OpenWhisk keeps functions hot for 
10 seconds and warm for 5 minutes, and the instances are reused based on this policy.
\ldfs{} always keeps function instances alive for the maximum allowed duration to
retain cached metadata in the function and does not consider \sysname{}-like dynamic function 
lifetime control. Each function instance of \ldfs{} runs for 5 minutes 
%(the default maximum function duration in OpenWhisk) 
and processes multiple concurrent user requests
during its lifetime. 
	
\begin{figure}
		\centering
		\includegraphics{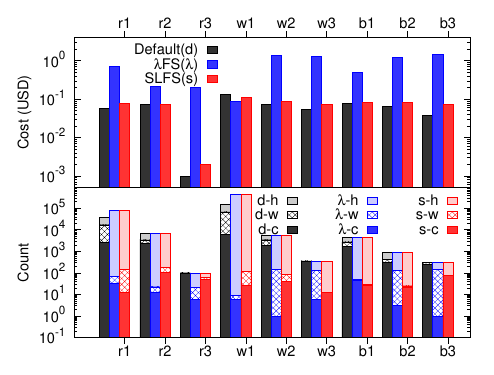}
	\vspace{-0.2in}
		\caption{Number of cold (c), warm (w), and hot (h) starts and function 
			invocation costs for default design (d), \ldfs{} ($\lambda$), and \sysname{} (s) in a \emph{log scale.}}
		\label{fig:cold-start} 
\end{figure}

\para{Frequency.} 
Figure~\ref{fig:cold-start} shows the result, where the y-axis displays the 
count in a log scale. The server design of \sysname{} functions avoids cold and 
even warm starts significantly compared to the default baseline: maximum 580$\times$ 
for ``w1'', on average 100$\times$, and on median 29.5$\times$. 
The Azure trace was collected on a commercial cloud-scale storage system, and
the default baseline could not keep up with the replay speed of some traces
in our cluster. The default design completed only 45.4\% and 31.8\% of the requests
for the short-interval ``r1'' and ``w1'' traces, respectively, whereas 
\sysname{} successfully completed all I/Os showing the effectiveness of the 
\sysname{} function design. Compared to \ldfs{}, which keeps the function instance
alive for the maximum allowed duration, \sysname{} exhibits a comparable number of 
hot starts: on average 3\% less than \ldfs{}. Note that the initialization time 
of a \ldfs{} function, even under warm start, takes much longer than the cold start 
time of the \sysname{} function instance to set up the long-running server,
establish connections with the persistent database, and initialize the 
in-memory database. Also, to keep \ldfs{} up to speed, we omit \ldfs{} data requests 
which do not use function instances and do not affect the cost measurement.

\para{Cost.} Comparing the monetary cost for function invocations, \sysname{} extends the function 
duration and incurs an extra cost for most cases compared to the default baseline: 
on average, the function invocation cost of \sysname{} is 33.9\% higher. If we project 
the cost to the full trace duration of 2 weeks, the maximum extra cost is only \$0.92.
This translates to an \sysname{} function running for, on average, 289 ms per I/O request
which is also reflected in our cost comparison in \cref{fig:awsprice}.
However, \ldfs{} which yields comparable hot start results to \sysname{} costs 
significantly more than \sysname{} as its function instance always runs until the maximum allowed duration: 
on average \ldfs{} costs 21 times higher than \sysname{}. 
This clearly shows the need and effectiveness of the function lifetime management of \sysname{}. 

\begin{figure}
	\centering
	\includegraphics{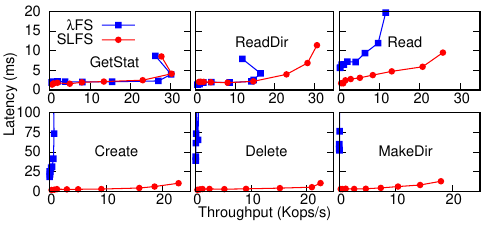}
	\vspace{-0.1in}
	\caption{Metadata read (top row) and metadata update (bottom row) operations of SLFS and \ldfs{}.}
	\label{fig:ldfs-read}
\end{figure}

\subsection{Metadata Operations under \ldfs{} Benchmark} 
\label{subsec:eval:meta}

We use \ldfs{} benchmark~\cite{lambdafsbench} to saturate the throughput to 
compare the performance of file system metadata operations of \ldfs{} and 
\sysname{} in the Amazon Cloud. 
%We increase the load until the throughput is saturated. 

For read-only metadata operations (Figure~\ref{fig:ldfs-read}, top row), \ldfs{} and 
\sysname{} perform similarly for GetStat, but \sysname{}
scales better for reading directories (ReadDir) and reading empty files (Read). 
Even though \ldfs{} function instances work as concurrent servers, finer-grained 
function instance management in \sysname{} makes operations more concurrent
and scalable.% than \ldfs{}.

For metadata operations involving updates (Figure~\ref{fig:ldfs-read}, bottom row), \sysname{}
significantly outperforms \ldfs{} for all operations: \ldfs{} scales up to only less 
than 1\,Kops/s, whereas \sysname{} scales to approximately 20\,Kops/s for all 
operations. One of the significant performance bottlenecks in \ldfs{} is the MySQL 
Cluster NDB for persisting metadata updates. \sysname{} is designed to work with 
simple key-value interfaces, which can scale better under the same hardware configuration.

\subsection{End-to-End Runtime for Azure Workloads}
\label{subsec:e2e}

\begin{figure}
    \centering
    \includegraphics{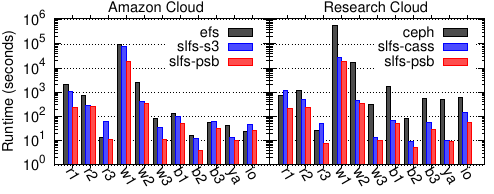}
	\vspace{-0.1in}
    \caption{End-to-end runtime comparison against EFS and Ceph in Amazon and public research clouds, respectively. 
	SLFS is configured to run on the PSB and cloud-native S3 or open-source Cassandra. The y-axis is
	in a log scale.}
    \label{fig:performance} 
\end{figure}

We compare the performance of \sysname{} against EFS~\cite{awsefs} and Ceph~\cite{ceph} 
in the Amazon and 
a public research cloud, respectively (Figure~\ref{fig:performance}, the 
y-axis is in a log scale).  We replay the Azure traces to their full length
as quickly as possible and run the YCSB-A workload and the IOZone file system 
benchmark (``ya'' and ``io'', respectively). IOZone  
issues concurrent 50 to 50 random reads and rewrites to large files.

\para{Comparisons with EFS.} In the Amazon cloud (\cref{fig:performance}-left), 
we run \sysname{} on S3 (standard) and PSBs (with gp2 EBS volumes) 
and against EFS (standard store, general purpose mode).
The experiment shows that \sysname{} on PSB outperforms EFS, on average, by 
4.27$\times$, 6.43$\times$, and 3.02$\times$ for read-only (r1-3), write-only (w1-3), and read-write 
workloads (b1-3, ya, and io), respectively. EFS is closed-source, but its 
documentation~\cite{efsslow} partly explains why it exhibits high overhead 
especially for write-intensive workloads. Accessing small files continuously adds 
cumulative latency and reduces the throughput. Additionally, small writes entail metadata 
modification and significantly degrade the performance 
(e.g., ``w1'' and ``w2''). \sysname{} avoids accessing block indexing data structures 
through hash-based indexes. As a result, \sysname{} outperforms EFS 
on average by 4.19$\times$.

Comparing \sysname{} on S3 and PSB, the PSB backend, which is more expensive, 
leads to 2.73$\times$ higher performance. 
Still, the cheapest \sysname{} configuration on S3 runs, on average, 1.99$\times$ faster than EFS.

\para{Comparisons with Ceph.} In a public research cloud (\cref{fig:performance}-right),
we conduct a similar experiment, where we run \sysname{} on open-source Cassandra and PSBs. 
The overall trend is similar to the experiment performed in the Amazon cloud and
\sysname{} on PSB exhibits very similar results to running in the Amazon cloud.
Ceph runs the slowest with larger gaps to \sysname{} when there are more writes: i.e.,
\sysname{} on PSB outperforms Ceph by 2.42$\times$, 16.98$\times$, and 23.73$\times$
for read-only, write-only, and mixed workloads respectively. 
A closer look at Ceph and the block drives provided by the research cloud reveals that 
the drives are very sensitive to random I/O as they are Cinder volumes backed by
another backend Ceph cluster and Ceph's architecture requires metadata updates 
through monitor and metadata server daemons for each file operation. 
While we collocated these daemons to minimize overhead, small random writes still 
require frequent coordination. In contrast, SLFS's hash-based 
indexing eliminates explicit metadata updates for most operations, avoiding 
this coordination overhead and PSB uses the log-structured LevelDB which is less 
vulnerable to small random writes. For similar reasons, Cassandra-based 
\sysname{} performs worse than PSB-based \sysname{}, but outperforms Ceph significantly. 

Despite the positive performance results from two different environments, 
we do not intend to argue strongly that \sysname{} is faster than EFS and Ceph considering the 
design differences and EFS and Ceph being stable production systems with more features. 
However, we claim that \sysname{} can be a scalable and cost-effective alternative to 
cloud-native or open-source file systems. Also, this experiment shows the flexibility 
of \sysname{} that can run on different backend stores with different cost and 
performance characteristics. 

\subsection{Cost Comparisons}
\label{subsec:eval:cost}
The unit cost of running serverless functions is generally more expensive than running a VM:
i.e., running a function instance non-stop for an hour is more expensive than renting 
a VM with the same amount of resources for an hour. For this reason, the cost benefit
of a serverless system becomes apparent when request intervals are reflected in the workload.
Therefore, experiments in \cref{subsec:eval:meta,subsec:e2e}, which feed in a large 
continuous stream of I/O workloads, demonstrate the peak performance of \sysname{},
but cannot represent a scenario for cost comparisons of running the system in production. 

\para{Comparisons with EFS and \ldfs{}.} Hence, we compare the cost of running EFS and \ldfs{} against \sysname{} based on the Azure
workload following the marked timestamps. 
\cref{fig:awsprice} considers all costs involved in maintaining the system while processing
user requests for one month under the same hardware configurations we used for the
experiments. The function invocation cost for \sysname{} is derived from
the experiments in \cref{subsec:eval:cold}. Depending on the configuration, \sysname{} can be
12\% to 68\% cheaper than EFS. While \sysname{} provides such flexibility, even the cheapest 
\sysname{} setup performs better than EFS (\cref{fig:performance}).

\cref{fig:cold-start} shows that the function invocation cost of \ldfs{} is 21 times 
higher than \sysname{} for representative workloads. Even considering other hardware costs, 
which can be the same as \sysname{} as in our experiment, the total cost of \ldfs{} becomes 
dominated by the function invocation cost. Analyzing the same way as in \cref{fig:awsprice}, the total cost of \ldfs{} 
becomes 31\% to 70\% more expensive (not in the figure but ranging from \$1,767 to \$2,289 depending on the EBS setting) 
than the EFS cost. Comparing matching EBS configurations, the total cost of \sysname{} is 48\% to 63\% cheaper than 
\ldfs{}, making \sysname{} the most cost-effective and performant option.

\para{Comparisons with Ceph and Cost per Performance.} Based on the configurations we choose under the same storage 
node setting (i.e., 9VMs) as \sysname{}, Ceph exhibits minimal hardware configuration and 
incurs constant hardware cost only for the storage VMs. Still, projecting the cost in \cref{fig:awsprice}
assuming the Amazon cloud deployment, Ceph becomes more expensive than \sysname{} on S3 even with 
the cheapest EBS volume type (i.e., Ceph costs \$537 on HDD-sc1 which is between SLFS-S3-std 
and SLFS-HDD-sc1 in \cref{fig:awsprice}). 

Since the Ceph price can be low, we further compare the price per work done based on the experiment 
in \cref{fig:performance} considering the compute cost (i.e., cost for functions and VMs) given
the same storage setting.  For \sysname{} on Cassandra, \sysname{} on PSB, and Ceph, the total price 
incurred based on the Amazon pricing model~\cite{awspricecalc} to run the Azure trace used in 
the figure is \$0.62, \$0.48, and \$1.31, respectively. Thus, Ceph is 2.09$\times$ to 2.72$\times$ more 
expensive than \sysname{} to do the same work, and this is due to lower performance and extended duration
to complete the work. Compared to Ceph, \sysname{} is more economical and 
provides more diverse cost configurations.

\section{Related Work}

\label{sec:related}
\para{Serverless In-Memory and Persistent Storage Systems.}
InfiniCache~\cite{infinicache} %is similar in spirit to \sysname{} as it
builds a distributed in-memory cache out of serverless function instances. 
\sysname{} uses the \proxy{} and InfiniCache uses the proxy to relay
the request to function instances. However, \sysname{} further extends
the role of the \proxy{} to make policy-based decisions for hot starts and 
better resource management. InfiniCache does not offer data durability and heavily relies on 
warm function containers over which InfiniCache has little control.
InfiniStore~\cite{infinistore} extends InfiniCache with
a serverless memory abstraction and persistence using a cloud object store 
to tolerate failures. However, it does not support file abstractions and
lacks the fine-grained function control of \sysname{}.

\ldfs{} augments HDFS-based HopsFS~\cite{hopsfs} with the InfiniCache idea 
to cache only metadata in a long-lived function instance. 
Its coordination service and name nodes resemble the 
\proxy{} and \Sdmfs{} of \sysname{} but \sysname{} is more general and 
supports features not in \ldfs{}. \sysname{} handles both file data and 
metadata through finely managed functions, dynamically controls function 
lifetime for performance and cost, and is portable to different backends.

\para{Cold Start and Function Lifetime Management.}
Cold start problems are largely tackled by
%There are two approaches to solve the cold start problem:
exploring lightweight mechanisms to start the function
instance~\cite{agache_firecracker_2020,catalyzer,ustiugov_benchmarking_2021, seuss}
and keeping function instances warm based on predicted request 
patterns~\cite{shahrad_serverless_2020, icebreaker, faascache}.
The \sysname{} solution is closer to the latter but extends the hot function 
state for a short controlled period to avoid both cold and warm starts; 
this is complementary to existing approaches and opens up new 
opportunities to even avoid warm starts. 

\para{Support for Stateful Serverless Computing.}
Pocket~\cite{pocket} is an ephemeral storage system for short-lived serverless 
tasks, Boki~\cite{boki} realizes a distributed shared 
log for serverless functions to coordinate, and Cloudburst~\cite{cloudburst} 
extends the auto-scaling key-value store Anna~\cite{anna} with
HydroCache~\cite{hydrocache} to collocate data and computation with a causal 
consistency guarantee. Similar to these systems, \sysname{} improves the 
accessibility of state across different serverless functions over persistent 
backend storage systems. Beldi~\cite{beldi} implements a serverless 
transactional key-value store with exactly-once semantics. Like Beldi, 
\sysname{} leverages the atomic commit features provided by the backend storage 
system and distributed transactions for consistent updates.
\sysname{} does not guarantee exactly-once semantics, which is beyond what 
most file systems provide, but Beldi's approach applies complementarily. 
Among all these systems, \sysname{} is the only file service 
leveraging serverless functions.

\para{Decomposed File Systems.}
There are serverful distributed~\cite{ceph, panasas} and local~\cite{kevinfs, kvfs, 
tablefs, betrefs} file systems designed on top of key-value 
stores. While local file systems leverage key-value stores for more
efficient indexing and fast metadata access, distributed file systems use 
them for modularity of design and ease of system management. 
FUSE-based file systems on cloud backend key-value stores~\cite{s3fs,gcsfs,agni}
are designed to leverage low-cost cloud key-value stores. 
While \sysname{} also uses key-value stores to delegate the complexity of fine-grained data 
management and to utilize low-cost cloud backend stores, \sysname{} goes
further to achieve utmost elasticity and even lower costs through serverless designs. 
Existing systems may be able to leverage container as a service at best without 
code modification for elasticity, but due to a heavier code footprint and larger
image size, their startup time would be nearly a 
minute~\cite{ec2start} to tens of seconds~\cite{fargatecoldstart1, 
fargatecoldstart2}, so they cannot be as elastic as \sysname{}.

CNFS~\cite{cnfs}, a cloud-native file system, advocates for designing storage systems around cloud primitives. CNFS is a local, hierarchical, copy-on-write file system that migrates data and metadata across cloud storage volumes and offloads background tasks such as migration and compression to remote CPU workers. While CNFS shares the high-level vision of decomposing file system functionality across cloud services, it focuses on local file system design principles and does not leverage the serverless computing model. \sysname{} is a fully distributed file system that uses serverless functions and proposes designs and methodologies to fully leverage the serverless environment.

\section{Conclusions}

We presented \sysname{}, the first distributed file system 
mainly designed with the serverless paradigm. \sysname{} can run on cloud-native 
and users' custom key-value stores and adjust to users'
needs. \sysname{} implements file operations using serverless functions. 
The function instances are designed as novel short-lived multi-threaded servers 
that dramatically avoid cold and warm starts at a low cost. \sysname{} makes 
policy-based decisions to scale the system elastically and uses residual 
function instance memory for caching. Our evaluation in the cloud with 
Azure workloads shows that \sysname{} performs better than 
\ldfs{}, EFS, and Ceph at a significantly lower cost.

%\begin{acks}
%\end{acks}

\bibliographystyle{ACM-Reference-Format}
\bibliography{refs}

\end{document}